# High Speed Elephant Flow Detection Under Partial Information

Jordi Ros-Giralt, Alan Commike
Reservoir Labs
632 Broadway Suite 803, New York, NY 10012

Sourav Maji, Malathi Veeraraghavan
Dept. of Electrical and Computer Eng., University of
Virginia Charlottesville, VA 22904–4743

*Abstract*—In this paper we introduce a new framework to detect elephant flows at very high speed rates and under uncertainty. The framework provides exact mathematical formulas to compute the detection likelihood and introduces a new flow Reconstruction Lemma under partial information. These theoretical results lead to the design of BubbleCache, a new elephant flow detection algorithm designed to operate near the optimal tradeoff between computational scalability and accuracy by dynamically tracking the traffic's natural cutoff sampling rate. We demonstrate on a real world 100 Gbps network that the BubbleCache algorithm helps reduce the computational cost by a factor of 1000 and the memory requirements by a factor of 100 while detecting the top largest flows on the network with very high probability.

## I. Introduction

A general objective in the design of high-performance computer networks is to guarantee the quality of service (QoS) experienced by the data flows that traverse them. This objective is often challenged by the presence of very large flows—also known as *elephant flows*—due to their adverse effects on smaller delay-sensitive flows. Because in these networks both large and small flows share common resources, network operators are interested in actively detecting elephant flows and using QoS mechanisms for redirecting and scheduling them to protect the smaller flows.

In this paper we focus on the problem of elephant flow detection at very high speed rates and under uncertainty. Sources of uncertainty can come from either a natural inability to predict the traffic's future performance or from artifacts introduced by networking equipment such as involuntary packet drops or voluntary packet sampling from protocols like sFlow [1]. The problem of identifying the minimum amount of information needed to detect the largest flows in a network is addressed. Then, under the assumption of heavy tailed traffic, we demonstrate the existence of cutoff sampling rates. Similar to the concept of *Nyquist sampling rate* in signal processing, the cutoff sampling rate of a traffic dataset corresponds to the minimum rate at which traffic must be sampled in order to detect and reconstruct the top largest flows with high probability.

Our theoretical framework provides two key building blocks for the design of optimal high performance elephant flow detection algorithms. First, it provides exact formulas to compute the detection likelihood, which reveal the necessary logic to ensure the algorithm targets an operational regime near the optimal tradeoff between computational scalability and accuracy. Second, the theory introduces the *Flow Reconstruction Lemma*, which states that if the sampled traffic dataset is heavy tailed, then the detection system operates error free with high probability. This lemma provides the necessary logic to ensure the convergence and stability of the detection algorithm.

We use the theoretic framework to design the *BubbleCache algorithm*, a high performance flow cache algorithm that captures the top largest (elephant) flows by dynamically tracking the optimal cutoff sampling rate inherent to the network traffic it processes. We demonstrate on a 100 Gbps network with real world IP traffic that the BubbleCache algorithm can help reduce the computational cost by a factor of 1000 and the memory requirements by a factor of 100 while detecting the largest flows on the network with high probability. Two direct applications of the BubbleCache algorithm are the design of optimal packet sampling modules such as those used in protocols like sFlow [1] and the design of high performance queues to dynamically separate elephant and mouse flows and to protect them from each other.

This paper is organized as follows. We omit a dedicated prior art section [2][3][4][5][6][7][8][9] to avoid redundancies as the description of the existing work in the literature is done throughout the body of the paper in direct comparison with our new results. Section II introduces the theoretical framework to detect elephant flows under partial information. This includes the detection likelihood equations, the Reconstruction Lemma and the resulting base algorithm to detect elephant flows by exploiting the traffic's natural cutoff sampling rates. In Section III we benchmark the performance of our proposed algorithm both in a controlled lab environment and in a real world 100 Gbps network. We conclude this paper in Section IV.

## II. Theory of Flow Ordering Under Partial Information

### A. On The Effect of Sampling

Consider a simple initial problem with a traffic dataset consisting of one single flow carrying m packets and *n* flows carrying 1 single packet. Fig. 1 displays the packet distribution corresponding to this traffic dataset.

Our interest is in finding a sampling strategy that allows us to identify the largest flow without necessarily processing all the traffic—that is, performing the detection under partial information. To resolve this problem, we observe that if we sample two packets from the elephant flow, then we can

This work was funded in part by the US Department of Energy under contract DE-SC0011358.

assert with certainty which flow is the biggest, since none of the other flows have more than 1 packet. In particular, let $X(k)$ be the number of packets sampled from the elephant flow out of a total of $k$ samples taken from the traffic dataset. Then the probability of identifying the elephant flow with certainty is:

$$P(X(k) \geq 2) = 1 - P(X(k) = 0) - P(X(k) = 1) \quad (1)$$

Using combinatorics and a few math derivations, it's easy to see that the equation ruling $P(X(k) \geq 2)$ corresponds to:

$$P(X(k) \geq 2) = \begin{cases} 1 - \frac{\binom{n}{k}}{\binom{m+n}{k}} - \frac{m\binom{n}{k-1}}{\binom{m+n}{k}}, & \text{if } 2 \leq k \leq n \\ 1 - \frac{m}{\binom{m+n}{n+1}}, & \text{if } k = n+1 \\ 1, & \text{if } n+1 < k \leq n+m \end{cases} \quad (2)$$

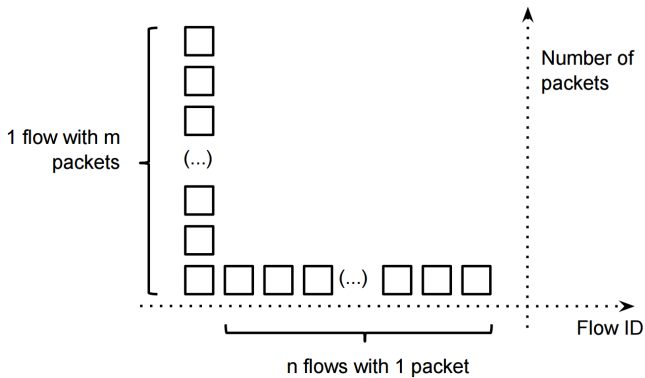

Fig. 1. A simple heavy-tailed traffic dataset

Fig. 2 plots the above equation for the case n = 1000, with $m$ varying from 1 to 15 and with $k = p \cdot (m+n)$, where $p$ is a sampling rate parameter between 0 and 1. We notice that:

- For the boundary case $m = 1$, the probability of finding the elephant flow is trivially zero, since the elephant flow is indistinguishable from the small flows.
- As we increase the sampling rate $p$, the probability of finding the elephant flow increases.
- As the number of packets in the elephant flow $m$ increases, we need less samples to gain a higher probability of finding it.

The intuition behind the previous result is as follows. Suppose that as observers of the network we see 10 packets from flow $f_1$ and 10 packets from flow $f_2$. We realize that we do not have enough information to make a good judgement as to which of the two flows is the largest. Suppose that instead, we see 100 packets from $f_1$ and 10 packets from $f_2$. If we had to make a guess, it seems reasonable to bet on $f_1$ being the largest of the two flows, but we may still not be convinced as we can't predict the future behavior of the two flows. Now suppose the case of seeing 1,000,000 packets from flow $f_1$ and only 10 packets from flow $f_2$. The chances of $f_2$ being the largest flow are now substantially lower, as it would need to transmit a very large number of packets to catch up with $f_1$. The logic of this reasoning is captured by Equation (2).

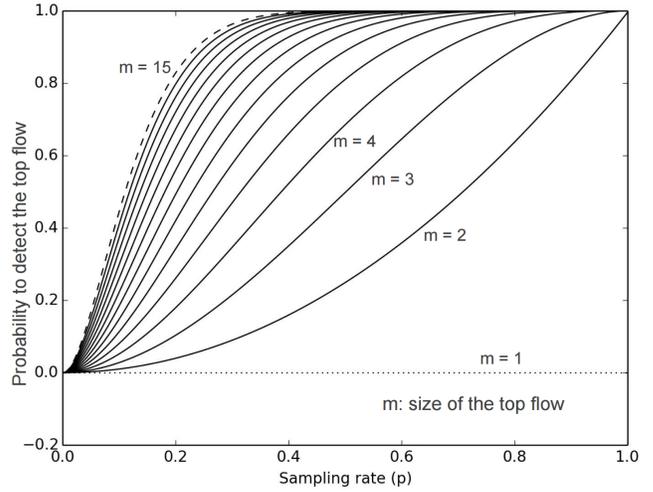

Fig. 2. Probability to detect the top flow

Another interpretation of Equation (2) in our simple network model is that it allows us to measure the likelihood of detecting the elephant flow as a function of uncertainty or the degree of partial information. When the sampling rate $p$ is 1, we have complete information and we can identify the elephant flow with certainty. As the sampling rate decreases to zero, the degree of partial information increases and the likelihood to detect the elephant flow decreases. In general, there are two sources of uncertainty that determine the effective sampling rate of our detection problem:

- *Future uncertainty.* Unlike oracles, we generally cannot predict the traffic that each flow will transmit in the future. To avoid this source of uncertainty, we need to wait until the last packet of all flows has been transmitted, but this is not practical since the objective of detecting elephant flows is to perform timely traffic engineering decisions while the flows are still active.
- *Past uncertainty.* Even if we could predict the future traffic transmitted by each flow, oftentimes networking equipment cannot keep up with the rates at which packets are processed in the data plane. For instance, in today's networks, it is computationally expensive to monitor every single packet going through a 100 Gbps link. Under these conditions, packets often need to be sampled or dropped, adding another source of uncertainty.

In the theoretical and algorithmic results presented in this paper, we will assume the network is under the influence of both of these sources of uncertainty.

Another interesting exercise is to contrast the implications of Equation (2) in our simple network model with the case of real world Internet traffic. It is well known that IP traffic is characterized by heavy tailedness [10][11], a condition in which traffic consists of a small number of flows transmitting a very large amount of data and a large number of flows transmitting a small amount of data. As illustrated in our simple example, this natural characteristic of Internet traffic works in favor of detecting the elephant flows with high likelihood under partial information: a larger value of *m*, implies a higher degree of heavy tailedness, which leads to a higher likelihood to detect the elephant flow. In Section II.D we mathematically formalize this concept with a lemma.

Hence our simple example in Fig. 2 offers some initial insights on the problem of elephant flow detection under partial information but its usefulness is limited in that it deals with a simple traffic dataset model consisting of 1 flow transmitting *m* packets and *n* flows transmitting 1 single packet. In the next section, we derive a generalized equation of the likelihood to detect elephant flows for arbitrary traffic distributions.

*B. Generalization to Arbitrary Distributions*

We start by introducing the definition of *quantum error* which will allow us to characterize the concept of detection likelihood for arbitrary traffic distributions:

*Definition 1. Quantum error (QER).* Let $F$ be a set of flows transmitting information over a network and let **x**(*t*) be a vector such that its $i$-th element, $x_i(t)$, corresponds to the size of flow $i$ at time $t$ according to some metric $m$. Examples of metrics can be the total number of bytes or packets transmitted by the flow [6], its average rate [8] or its burstiness [10], among many others. **x**(*t*) is therefore a time-varying vector such that $x_i(t_b) = 0$ and $x_i(t_e) = \sigma_i$, where $t_b$ and $t_e$ are the times at which the first (beginning time) and the last (ending time) bit of information are transmitted from any of the flows, and $\sigma_i$ is the size of flow $i$ at time $t_e$. Assume without loss of generality that $\sigma_i \geq \sigma_{i+1}$ and let $F_\alpha = \{f_1, f_2, ..., f_\alpha\}$ be the set with the $\alpha$ largest flows according to their size at time $t_e$, $x_i(t_e) = \sigma_i$, for some $\alpha \leq |F|$. Finally, let $C_\alpha(t)$ be a cache storing the top $\alpha$ largest flows according to their size at time $t$, $x_i(t)$. (Hence, by construction, $C_\alpha(t_e) = F_\alpha$.) We define the *quantum error* (*QER*) produced by the cache at time $t$ as:

$$e_\alpha(t) = \frac{|F_\alpha \setminus C_\alpha(t)|}{\alpha} = \\ = \frac{|\{x_i(t) \text{ s.t. } \sigma_i \leq \sigma_\alpha \text{ and } x_i(t) > x_\alpha(t)\}|}{\alpha} \quad (3)$$

§

Intuitively, the above equation corresponds to the number of small flows that at time $t$ are incorrectly classified as top flows normalized so that the error is 1 if all top $\alpha$ flows are misclassified. Because this error refers to the notion of an observer classifying a flow at an incorrect size order or level, we use the term *quantum* error or *QER*. We can now formally introduce the concept of detection likelihood:

*Definition 2. Top flow detection likelihood.* The *top flow detection likelihood* of a network at time $t$ is defined as the probability that the quantum error is zero: $P(e_\alpha(t) = 0)$. When the meaning is obvious, we will refer to this value simply as the *detection likelihood*. §

Using the above definition, we can mathematically derive the detection likelihood equation:

*Lemma 1. Detection under partial information.* The detection likelihood of a network at time $t$ follows a multivariate hypergeometric distribution as follows:

$$P(e_\alpha(t) = 0) = P(C_\alpha(t) = F_\alpha) = \sum_{\forall \mathbf{x}' \in Z(t)} \frac{\prod_{\forall i} \binom{\sigma_i}{x'_i}}{\binom{\sum_{\forall i} \sigma_i}{\sum_{\forall i} x_i(t)}} \quad (4)$$

where $Z(t)$ is the zero quantum error region, expressed as:

$$Z_\alpha(t) = \{\mathbf{x}' \in \mathbb{N}^{|F|} \mid \sum_{\forall i} x'_i = \sum_{\forall i} x_i(t), \\ \mathbf{x}' \leq_p \boldsymbol{\sigma}, \ x'_i > x'_j \ \forall i,j \text{ s.t. } i \leq \alpha, \ j > \alpha\} \quad (5)$$

and $a \leq_p b$ means $b$ is at least as *Pareto efficient* as $a$.

*Proof.* See Appendix.

§

As a test of generality, we can mathematically show that Equation (4) is a generalization of Equation (2) for arbitrary traffic distributions:

*Corollary 1. Test of generality.* Assume a traffic dataset consisting of one single flow carrying $m$ packets and $n$ flows carrying 1 single packet. Then the detection likelihood function presented in Equations (4) and (5) is equivalent to Equation (2).

*Proof.* See Appendix.

§

In the next section, we study the practical implications of Lemma 1 towards the design of high performance elephant flow detection algorithms.

*C. On the Minimum Information Needed to Detect Elephant Flows: Cutoff Sampling Rates*

From a practical standpoint, the detection likelihood $P(e_\alpha(t))$ in Equation (4) cannot be computed for times $t < t_e$ because the size of all flows $\sigma_i$ is only known with certainty at time $t = t_e$. Nevertheless, its equation reveals important properties related to the problem of elephant flow detection. Suppose a network switch inspects packets in real time with the goal of timely identifying the top largest flows, where a flow's size is determined by an arbitrary metric—e.g., packet counts, byte counts, rate, etc. Assume that, due to limitations in both computing power and memory footprint, the switch can only store in the cache a maximum of $\alpha$ flows. Then, the following statements about the detection likelihood equation are true:

- It provides the minimum amount of samples we need to inspect (equivalently, the minimum amount of time we

- need to wait) to make a classification decision that will be correct with a probability given by $P(e_\alpha(t) = 0)$ or higher.
- It mathematically quantifies the trade-off between time and the quantum error: if we trade time by waiting longer to make a detection decision, we can reduce quantum error; if we trade quantum error, we can make a detection decision sooner.

From an information theory standpoint, a relevant question is to identify the minimum amount of information that needs to be sampled from the traffic dataset in order to detect the largest flows for a given detection likelihood. This problem is similar to the concept of *Nyquist rate* in the field of signal processing, which identifies the minimum number of samples that need to be taken from a signal in order to fully reconstruct it [15]. We explore this problem in more detail through an example.

*Example 1. Minimum sampling rate of some well-known heavy tailed traffic distributions.* Let $F$ be the set of flows in a network and let $\sigma_i$ be the size of each flow $i$, for $1 \le i \le |F|$. Assume $\sigma_i$ follows any of these well-known distribution functions:

| Laplace | Cauchy | Sech-squared | Gaussian | Linear |
|---|---|---|---|---|
| $\sigma_i = \gamma \frac{1}{2} e^{-|i|}$ | $\sigma_i = \gamma \frac{1}{\pi(1+i^2)}$ | $\sigma_i = \gamma \frac{e^{-i}}{(1+e^{-i})^2}$ | $\sigma_i = \gamma \frac{e^{-i^2/2}}{\sqrt{2\pi}}$ | $\sigma_i = \gamma(|F| - i)$ |

where $\gamma$ is chosen so that $\sum_{\forall i} \sigma_i$ is a constant. Fig. 3 plots the detection likelihood using Equation (4) for the case that $\sum_{\forall i} \sigma_i = 300$, $\alpha = 5$ and $|F| = 40$ when a fraction $p$ of the traffic is sampled, for $0 \le p \le 1$. The cutoff rates that result in a detection likelihood of 0.99 are also computed. As expected, for non-heavy tailed traffic patterns such as the linear distribution, the cutoff rate is high at $p=0.97$, while the cutoff rate for heavy tailed patterns such as the Gaussian distribution is much lower at $p=0.01$. For instance, under the special case where the flow size metric corresponds to the number of packets in a flow, this means that for the Gaussian, Laplace, Sech-squared and Cauchy distributions it is enough to sample 1%, 3%, 7%, and 12% of the total traffic dataset, respectively, in order to detect the 5 largest flows with a 99% chance of being correct. §

It is also worth noting in Fig. 3 that a small reduction of the sampling rate below its cutoff rate results in a substantial reduction of the detection likelihood. This property leads to significant optimization opportunities in the design of high performance elephant flow detection algorithms. Consider as an example the Laplace distribution. Reducing the sampling rate from 1 to 0.03 results in practically no detection penalty, but it leads to computational savings of about 97% or, equivalently, a computational acceleration of 33 times. These cutoff rates, which depend only on the statistical properties of the traffic, define optimal operational regimes that are key to the design of computationally efficient detection algorithms as we will see later in our work.

Of interest is the problem of identifying the actual cutoff rates of real world Internet traffic towards identifying optimal packet sampling strategies. Later in Section III.B.1 we carry out this exercise by measuring the cutoff rates for a mix of live public and science traffic from a 100Gbps data network.

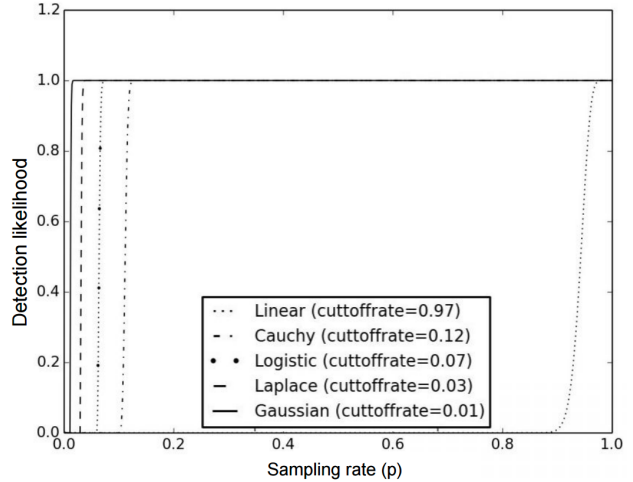

Fig 3. Detection likelihood of some known distributions

*D. High-Performance Detection Algorithms*

*1) Base Algorithm: The BubbleCache*

A good amount of elephant flow detection algorithms from the literature use packet sampling as a strategy to reduce computational complexity [5][6][7][8][9]. For instance, Psounis et al. [5] introduce an elegant low-complexity scheduler which relies on packet sampling to detect when a flow traversing a network switch is likely to be an elephant flow. In [6], the idea of packet sampling is generalized to design an actual *elephant trap*, a data structure that can efficiently retain the elephant flows and evict the mouse flows requiring low memory resources. These existing algorithms, however, treat the packet sampling rate as an input that operators need to manually adjust. Instead, our framework leads to an unmanned packet sampling algorithm that can dynamically adjust the sampling rate towards tracking a detection likelihood target. To the best of our knowledge, the algorithm we present is the first to exploit the concept of cutoff rates found in network traffic to compute the sampling rate of the detection algorithm and optimize the tradeoff between computational scalability and accuracy. Because of its generality, instead of a competing solution, the algorithm we present next can be used to enhance the existing packet sampling based elephant flow detection algorithms.

We know that heavy tailed traffic characteristics such as those found in real world networks expose detection likelihood curves with well defined cutoff rates, as illustrated in Fig. 3. Above the cutoff rate, the gains on the probability to accurately detect the largest flows are small. Below it, the penalties are large. A detection algorithm can benefit from this property by tuning its sampling rate to target the cutoff rate, substantially reducing the computational cost of processing traffic while controlling a small or negligible error rate. This suggests the following simple base algorithm to detect elephant flows at high speed traffic rates:

*Pseudocode 1: The base BubbleCache algorithm*

*Algorithm BubbleCache*
$\Phi$ : *Targeted accuracy parameter*
$\delta_p$ : *Sampling rate step size*
$T_i$ : *Inactivity timeout*
$T_h$ : *Housekeeping routine timeout*
$t$ : *The current time*
$C_\alpha(t)$ : *The state of the flow cache at time t*
$x_i(t)$ : *The measured size of flow i*
*m: flow size metric (e.g., bytecount, packet count, ...)*
*Upon receiving a packet from an arbitrary flow $f_i$:*
   *Sample the packet with a probability $p(t)$;*
   *If the packet is sampled:*
      *If the packet's flow is not in $C_\alpha(t)$:*
         *Add a new flow record to $C_\alpha(t)$ for the packet's flow;*
      *Update $x_i(t)$ according to the size metric m;*
*Every $T_h$ units of time:*
   *If undersampling(), increase $p(t)$ by $\delta_p$;*
   *Otherwise, reduce $p(t)$ by $\delta_p$;*
   *Remove flows from $C_\alpha(t)$ that have been inactive for $T_i$;*
*Function undersampling():*
   *If $P(e_\alpha(t) = 0)$ is lower than $\Phi$, return true;*
   *Else, return False;*

---

The central idea of the above pseudocode, referred as the *BubbleCache algorithm*, is to sample packets at a rate $p(t)$ which is updated to track a target detection likelihood: if the current detection likelihood $P(e_\alpha(t) = 0)$ is lower than a target $\Phi$, then $p(t)$ is increased; otherwise, $p(t)$ is decreased.

A practical limitation of the BubbleTrap algorithm is the calculation of the detection likelihood value $P(e_\alpha(t) = 0)$, as this formula implicitly assumes an oracle view of the network. In particular, Equation (4) requires the knowledge of the size of each flow, $\sigma_i$. But this implies full knowledge of the network state, not just from the past but also its future state, since $\sigma_i$ is the size of flow $f_i$ once all of its data has been transmitted, $\sigma_i = x_i(t_e)$. Using the law of large numbers, we can be overcome this assumption by providing an estimation of this value as follows:

*Corollary 2. Estimated detection likelihood.* Let an elephant flow detection algorithm process traffic by sampling packets at a rate $p$. Then, an estimation of the detection likelihood can be obtained as follows:

$$\hat{P}(e_\alpha(t) = 0) = \sum_{\forall x' \in \hat{Z}(t)} \frac{\prod_{\forall i} \binom{x_i(t)/p}{x'_i}}{\binom{\sum_{\forall i} x_i(t)/p}{\sum_{\forall i} x_i(t)}} \quad (8)$$

where $\hat{Z}(t)$ is the estimated zero quantum error:

$$\hat{Z}_\alpha(t) = \{x' \in \mathbb{N}^{|F|} \mid \sum_{\forall i} x'_i = \sum_{\forall i} x_i, \\ x' \leq_p x(t)/p, \; x'_i > x'_j \; \forall i,j \text{ s.t. } i \leq \alpha, j > \alpha\} \quad (9)$$

*Proof.* See Appendix.

§

While Equation (8) is now computable, there are still two challenges that make it impractical from an engineering perspective. First, in the BubbleCache algorithm, the sampling rate changes with time, which invalidates the assumption that $\sigma_i$ can be estimated with the expression $x_i(t)/p$, for a fixed $p$. While this issue can be solved by adjusting the equation to take into account integral changes in the sampling rate, a second more limiting issue arises due to its computational cost: as Equation (8) requires combinatorial operations, no current modern computer can calculate its value without overflowing the computation. In the next section we will elaborate a method to overcome both of these limitations.

*2) Estimating Detection Likelihoods*

In order to develop a computationally feasible approach to calculate detection likelihoods, we need to first formalize the definition of heavy tailed traffic and introduce the main reconstruction lemma on which our approach will be based:

*Definition 3. Heavy tailed traffic.* Let $F$ be a set of flows transmitting data over a network and assume $\sigma_i$ corresponds to the size of flow $i$ according to some metric $m$. Let also $F_e$ and $F_m$ be the set of elephant and mouse flows in $F$ according to this metric, respectively. We will say that the traffic dataset generated by the flows in $F$ is *heavy tailed* if $|F_e| << |F_m|$ and $\sigma_i >> \sigma_j$ for any pair of flows $f_i$ and $f_j$ in $F_e$ and $F_m$, respectively.

§

We now state the reconstruction lemma which will provide the blueprints of our proposed top flow detection algorithm:

*Lemma 2. Reconstruction under partial information.* Let $F$ be a set of flows transmitting data over a network and let $x_i$ be the size of flow $f_i$ when traffic is sampled at a rate $p$, for $0 \leq p \leq 1$. Then there exists a sampling rate $p_c < 1$ such that:

- If $x_i \gg x_j$ and $p \geq p_c$, then $\sigma_i \gg \sigma_j$ with high probability.
- If $\sigma_i \gg \sigma_j$ and $p \geq p_c$, then $x_i \gg x_j$ with high probability.

*Proof.* See Appendix.

§

Lemma 2 provides two interpretations of the elephant flow reconstruction problem which correspond to two faces of the same coin:

- If the measured signal $\{x_i\}$ is heavy tailed and if the sampling rate $p$ is high enough, with high probability the original signal $\{\sigma_i\}$ is also heavy tailed.
- If the original signal $\{\sigma_i\}$ is heavy tailed and if the sampling rate $p$ is high enough, with high probability the measured signal $\{x_i\}$ is also heavy tailed.

The concept of the "sampling rate $p$ being high enough" is characterized by the existence of a cutoff rate, $p_c$, above which the reconstruction Lemma holds. In Fig. 3, we had already seen the presence of these cutoff rates for a few heavy tailed traffic distributions: above the cutoff sampling rate, the detection likelihood is very high; but shifting the sampling rate slightly below it, the detection likelihood becomes very low. Lemma 2 should be interpreted as a mathematical statement of the existence of these cutoff sampling rates.

From the reconstruction Lemma, we can derive the following corollary which we will use to design our elephant flow detection algorithm:

*Corollary 3. Reconstruction properties under partial information.* Let $F$ be a set of flows transmitting data over a network and assume the traffic dataset generated by the flows is heavy tailed according to Definition 3. Let also $x_i$ be the size of flow $f_i$ when traffic is sampled at a rate $p$, for $0 \le p \le 1$ and $1 \le i \le |F|$. Then the following is true:

(R1) There exists a cutoff sampling rate $p_c$ such that for any sampling rate $p \ge p_c$, $\sigma_i \gg \sigma_j$ implies $x_i \gg x_j$ with high probability.

(R2) The more heavy tailed the traffic data set is, the lower the cutoff sampling rate $p_c$.

(R3) If the sequence $\{x_1, x_2, ..., x_{|F|}\}$ is heavy tailed, then $x_i \gg x_j$ implies $\sigma_i \gg \sigma_j$ with high probability.

(R4) If the sequence $\{x_1, x_2, ..., x_{|F|}\}$ is not heavy tailed, then either $p < p_c$ or the traffic dataset is not heavy tailed, or both.

*Proof.* See Appendix.

§

The Reconstruction Lemma and its Corollay have practical implications in the design of high performance algorithms to detect elephant flows. In particular, from Corollary 3/R4, if $\{x_1, x_2, ..., x_{|F|}\}$ is not heavy tailed, then either the traffic has no elephant flows or the sampling rate is too small, $p < p_c$. Assuming real world network traffic is heavy tailed (otherwise there would be no need to identify elephant flows to optimize network traffic), we can conclude that $p < p_c$ and hence that the sampling rate needs to be increased. If instead $\{x_1, x_2, ..., x_{|F|}\}$ is heavy tailed, then using Corollary 3/R3 we know that $x_i \gg x_j$ implies $\sigma_i \gg \sigma_j$ with high probability, and hence that the elephant flows can be clearly separated from the mouse flows by measuring $\{x_1, x_2, ..., x_{|F|}\}$ without the need to know the actual sizes of the flows $\{\sigma_1, \sigma_2, ..., \sigma_{|F|}\}$.

This reduces the hard problem of computing the detection likelihood $P(e_\alpha(t) = 0)$ to the simpler problem of measuring whether the input signal (the traffic under measurement) is heavy tailed: if the measured traffic is heavy tailed, then $p \ge p_c$ and we can identify the elephant flows with high probability. If the measured traffic is not heavy tailed, then we need to increase the sampling rate until it becomes heavy tailed.

To address this new objective, we propose to use the fourth standardized moment, known also as the *kurtosis* [16], which is simple to measure and provides the degree to which a signal is heavy tailed. In particular, the kurtosis of a sequence $\{x_1, x_2, ..., x_{|F|}\}$ can be computed as follows:

$$Kurt[\{x_1, x_2, ..., x_{|F|}\}] = \frac{\sum_{\forall i}(x_i - \bar{x})^4/|F|}{(\sum_{\forall i}(x_i - \bar{x})^2/|F|)^2} \quad (14)$$

Statistical moments are one of the most widely used tools in descriptive statistics and algorithms have been developed to efficiently compute them in the context of high performance computing. In [16], the author derives formulas for the real-time computation of arbitrary statistical moments. These formulas can be used to efficiently compute the value of Equation (14) as a series of $O(1)$ incremental updates performed every time a sample of the traffic dataset is processed.

To understand the intuition behind this approach consider Table 1, which presents the kurtosis of the traffic data sets introduced in Example 1. As expected, the four heavy tailed data sets (Laplace, Cauchy, Sech-squared and Gaussian distributions) present a high kurtosis (above 12), whereas the non-heavy tailed distribution (linear distribution) exposes a low kurtosis (-1.2). By using the kurtosis measurement, we can determine whether the sampled traffic dataset is heavy tailed and therefore if the detection likelihood is high according to Corollary 3/R3.

Table 1. Kurtosis of the traffic distributions in Example 1

| Linear | Laplace | Cauchy | Sech-squared | Gaussian |
|---|---|---|---|---|
| -1.2 | 25.88 | 20.54 | 12.11 | 18.86 |

The next pseudocode provides the simple adjustment needed on the base algorithm to enable the calculation of the cutoff sampling rate based on the kurtosis method:

*Pseudocode 2: undersampling() function with kurtosis*

*Function undersampling():*

*If $Kurt[\{x_1, x_2, ..., x_{|F|}\}]$ is lower than $\Phi$:*

 *Return true;*

*Else:*

 *Return False;*

As shown, the BubbleCache algorithm is remarkably simple: track the kurtosis metric of your input traffic and increase/decrease the sampling rate depending on whether the kurtosis value is below/above a target value. In the next section we provide detailed implementation and benchmarks demonstrating the robustness, simplicity and scalability of the BubbleCache algorithm.

## III. Performance Benchmarks

We have implemented the BubbleCache as a passive tapping network device—i.e., a device that processes a mirrored copy of the traffic without affecting any of the active networking equipment (routers, switches, hosts, etc.). The device specifications include two Intel Xeon E5-2670 processors clocked at 2.50 GHz for a total of 20 physical cores with 25.6MB of L3 cache for each processor. It also incorporates four 40 Gbps Solarflare SFC9100 SFP optical interfaces steered by DNAC [17], a high performance packet forwarding engine that performs line rate per-flow load balancing from the network ports to the processor cores. Each core is programmed to run a replica of the BubbleCache algorithm presented in Pseudocode 1 with the *undersampling()* method based on the kurtosis measurement as described in Pseudocode 2. Throughout the benchmarks, we assume a flow size metric $m$ corresponding to the number of packets in a flow (see Definition 1).

We performed two sets of benchmarks. First, we measured the performance of the BubbleCache under a controlled lab environment. For these tests, the sampling rate was statically set, which allowed us to make fine-grained measurements of the quantum error at various sampling rates. The second set of benchmarks consisted of a series of high-performance tests carried out while running the BubbleCache device live at last year's SuperComputing (SC) Conference.

### A. Measurements with Static Sampling

In this section we present the results of testing the BubbleCache device in a controlled lab environment using traffic from our corporation's local area network. This traffic, which we will refer as the LAN traffic dataset, includes a mix of machine generated flows (for services such as SNMP) and human generated traffic (for applications such as HTTP/HTTPS). The high level statistics of the packet trace are described in table 2. Because our goal is to measure the performance of the BubbleCache at fixed sampling rates, for the tests in this section we modify Pseudocode 1 to keep the sampling rate constant at a predetermined value of our choice.

Table 2. Statistics of the LAN traffic dataset

| TCP | UDP | ICMP | Other | Avg pkt size | Size |
|---|---|---|---|---|---|
| 96.74% | 3.17% | 0.03% | 0.06% | 592.61 | 120GB |
| **DHCP** | **MS-DS** | **HTTP(S)** | **SMTP** | **SSH** | **Other** |
| 0.05% | 2.95% | 4.91% | 0.65% | 49.66% | 41.78% |

*1) Cutoff Sampling Rate*

We start our tests by measuring the natural cutoff sampling rate of the LAN traffic. To do this measurement, we replay the trace at the rate it was captured to ensure there is no packet loss and measure the quantum error due to sampling by using Equation (3). The results are illustrated in Fig. 4. We note that the LAN dataset accepts a sampling rate of $p = 0.05$ while maintaining a zero quantum error (QER).

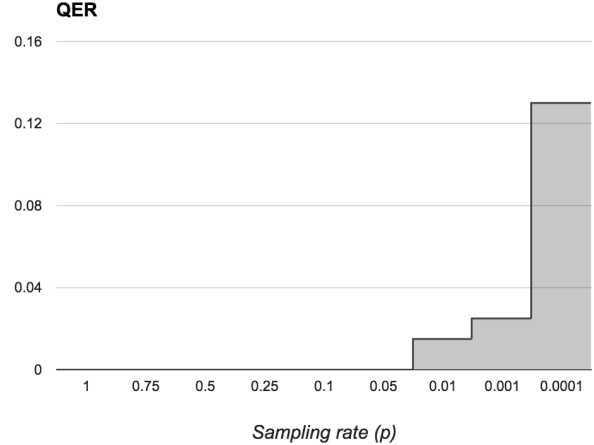

Fig. 4. Cutoff sampling rate of the LAN traffic dataset

*2) Optimal Sampling Rate at 100 Gbps*

In the next experiment, we replay the LAN dataset at a rate of 100 Gbps and measure the QER of the BubbleCache as a function of the sampling rate. The results in Fig. 5 illustrate an expected U-shape with three different regions. For very low sampling rates (p=0.001 and below), the QER rapidly increases due to excessive sampling. For high sampling rates (p=0.9 and above) the QER is also high due to the BubbleCache device not being able to keep up with the 100 Gbps traffic rates, resulting in packet drops. The optimal sampling rate sits somewhere between these two edge cases, at around p=0.46 resulting in a QER value of 0.00875.

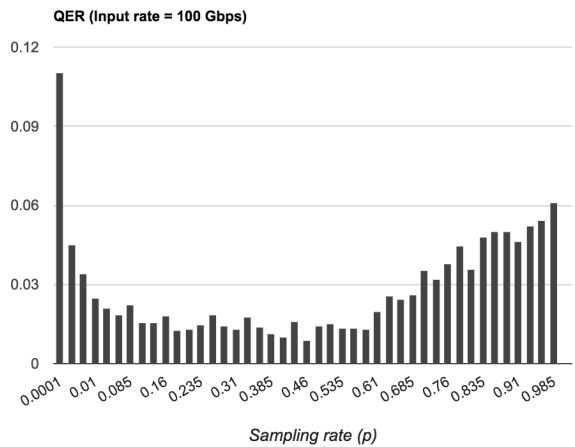

Fig 5. QER at 100 Gbps as a function of sampling

## 3) Quantum Error and Packet Drops

Fig. 6 presents quantum errors and packet drops at different traffic and sampling rates. The results indicate that using no sampling ($p=1$) is sub-optimal for rates of 77 Gbps and above, in the region where the QER becomes larger than zero (Fig. 6-a). For this region, the QER can be reduced by progressively incrementing the sampling rate to the neighborhood of $p=0.4$ (Fig. 6-b), in agreement with the results in Fig. 5. Increasing the sampling rate beyond this value (Fig. 6-c) helps further reduce packet drops but it has a negative effect on the QER.

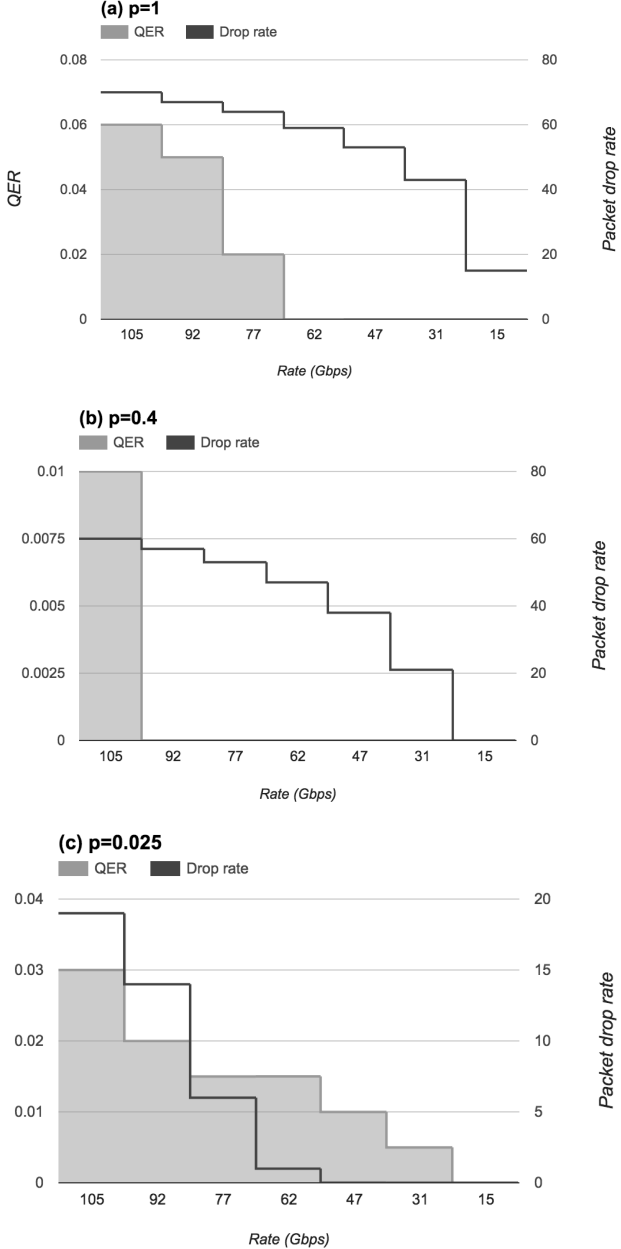

Fig. 6. Quantum errors and packet drops

## B. Measurements with Dynamic Sampling

We tested the dynamic version of the BubbleCache algorithm in a live high performance network environment with the goal to: (1) measure the natural cutoff sampling rate of traffic from a real world IP network and its variations throughout time, (2) measure the convergence and stability of the dynamic sampling rate algorithm and (3) measure the computational and memory footprint savings obtained by operating at the neighborhood of the cutoff sampling rate. These experiments were performed during the days of November 14 through 18, at the SuperComputing (SC) venue as part of the high performance computing (HPC) demonstrations run in the SCinet network. The SCinet HPC network is built every year to help support the SC venue and to test new technologies in a realistic network environment. This large scale network environment supports a traffic mix of both small flows generated by thousands of users on the conference floor and very large flows generated by large scale, big data science experiments carried out from the booths (where many of the US National and International Labs, Universities and companies carry their high speed experiments), resembling the traffic conditions typically found in Research and Education networks such as ESnet and Internet2.

As part of the SC/SCinet team, we connected the 4x40 Gbps ports of the BubbleCache device to one of the network taps which had full visibility of the SC/SCinet traffic. For these tests, the BubbleCache algorithm was configured with the following parameters: $\Phi = 100$ (target kurtosis value), $\delta_p = 0.01$ (sampling rate step size), $T_i = 20$ seconds (connection inactivity timeout), $T_h = 0.05$ seconds (housekeeping routine timeout). The rationale for choosing a target kurtosis value of 100 is to conservatively operate the algorithm at a region where the quantum error is zero with very high probability. Notice that heavy tailed functions such as those presented in Examples 5 and 6 (Laplace, Cauchy, Sech-squared and Gaussian distributions) have kurtosis values between 10 and 25; hence, a value of 100 ensures that the sampled traffic dataset is very heavy tailed. From Lemma 2, this in turn implies the algorithm operates at the zero quantum error region with high probability. To test the efficacy of the BubbleCache device under limited computing resources, we configured the device to only use four cores out of its total 20 cores, with each core processing one of the four 40 Gbps network ports, while leaving the other 16 cores idle.

## 1) Cutoff Sampling Rate for IP Traffic and Convergence Measurements

While the existence of cutoff sampling rates was mathematically shown in Lemma 2 and illustrated in Fig. 3, a question of interest is whether their presence can also be measured on real live traffic. In particular, we are interested in answering: what is the cutoff sampling rate of a real world

IP network? and how does this cutoff rate change as traffic patterns in the network change throughout the day?

Fig. 7 presents the sampling rate obtained from running the BubbleCache algorithm for traffic generated from the SC/SCinet network during high and low traffic hours. The traffic rate coming from the venue floor during high hours (during the day) was around 25 Gbps with peaks at 60 Gbps, whereas at low traffic hours (at night) traffic was around 1Gbps or below. With a target kurtosis of 100, the cutoff sampling rate at high and low traffic hours is around 0.001 and 0.01, respectively. This result shows that at traffic rates of about 25 Gbps, we can sample around 1 out of 1000 packets (a computational cost reduction of 1000 times) and still capture all the largest flows with high probability as the resulting sampled traffic dataset is very heavy tailed. Another result worth noting is that the higher the traffic rates, the lower we can reduce the sampling rate for a fixed target kurtosis level (i.e., a fixed degree of heavy tailedness). Using Corollary 3/R2, this implies network traffic must be more heavy tailed during the day, which is in agreement with the fact that more heavy tailed traffic is produced during the conference hours when the big data science experiments launched throughout the day are combined with thousands of user-generated small flows. This result is relevant in that it is at very high speed rates that a reduction of the sampling rate becomes most valuable from a computational scalability point of view. The BubbleCache algorithm is able to leverage this natural property of the traffic by tuning the sampling rate up or down as necessary.

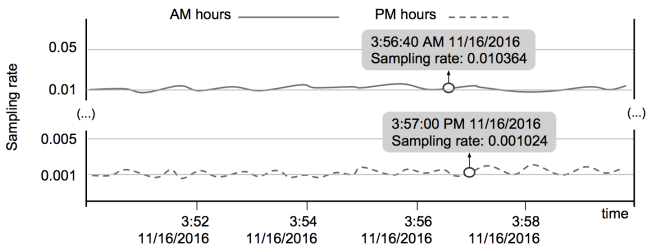

Fig. 7. Measurements of the cutoff sampling rate

Fig. 8 plots the convergence of both the sampling rate and the kurtosis parameters as the algorithm is started from two different initial conditions during the high traffic hours (around 2:10pm). In Fig. 8-top, the initial sampling rate is set to 0.0001, ten times below the optimal rate of 0.001, while in Fig. 8-bottom, the initial sampling rate is set to 0.01, ten times above it. In both cases, in a few seconds the algorithm converges to the same cutoff sampling rate around 0.001. The convergence time is linear and its slope can be tuned by adjusting the sampling rate step size $\delta_p$ and the housekeeping routine timeout $T_h$ (see Pseudocode 1). While left outside the scope of these results, an area of optimization is to improve the convergence time by using an adaptive heuristic that increases the step size if the kurtosis index is far from the target and reduces the step size as it gets closer to it.

In summary, the above plots show that, regardless of the initial conditions, the sampling rate converges to the targeted kurtosis value of 100 and, upon convergence, both the sampling rate and the kurtosis parameters stay stable around their targets.

*2) Memory Footprint*

In addition to the computational savings shown in the previous section, sampling also has a positive effect on the memory footprint requirements of the algorithm: the higher the sampling rate, the smaller the size of the flow cache as more flows are filtered out. We are now interested in measuring the memory footprint reduction accomplished as a consequence of sampling traffic at the targeted kurtosis level.

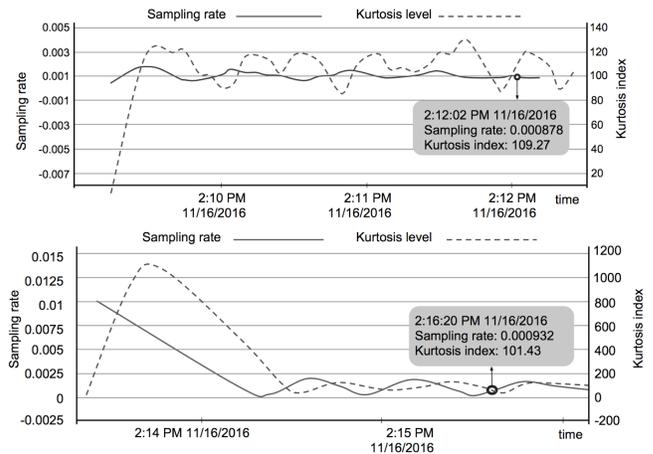

Fig. 8. Convergence of the BubbleCache algorithm

Fig. 9 illustrates the size of the BubbleCache as a function of time as the algorithm converges to the cutoff rate of 0.001 from an initial sampling rate of 0.01. (This plot captures the same time period as the plot in Fig. 8-bottom.) At any point in time, the average number of active flows in the SC/SCinet network for this period is around 25,000. As the BubbleCache algorithm is initiated, since the sampling rate is substantially above the cutoff rate, the size of the flow cache steadily increases reaching more than 2000 flow entries. Then as the sampling rate and the kurtosis level continue to decrease, the size of the cache begins to decrease until it reaches a stable point once the targeted kurtosis level of 100 is achieved. In steady state and with 25,000 active flows, the size of the flow cache stabilizes around 250 flows, which represents a 100 time reduction in memory size.

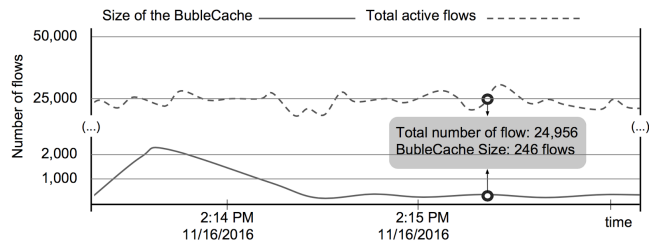

Fig. 9. Size of the BubbleCache

## 2) Flow Cache Heavy Tailedness

From Corollary 3/R3, we know that if the sampled distribution is heavy tailed, then with high probability the top largest flows are captured in the cache. Intuitively, as the flow cache is capable of capturing the heavy tailedness of the input traffic, the probability of quantum error tends to zero. To validate this concept, Fig. 10 plots the average sizes of the flows captured by the BubbleCache during high (day time) and low (night time) traffic hours on a log scale graph. As shown, the BubbleCache dynamically adjusts its size to ensure the captured flows expose the heavy tailed shape. From Lemma 2, since the BubbleCache captures flows that are much smaller than other flows, $x_i \gg x_j$ for some flows $f_i$ and $f_j$, then we can conclude that with high probability all the largest flows are captured and the quantum error is zero. If the sampling rate were reduced further below the cutoff rates (0.01 at night time and 0.001 at day time), then the kurtosis would also decrease, eventually eliminating the heavy tailed shape of the sampled traffic and increasing the likelihood of a quantum error.

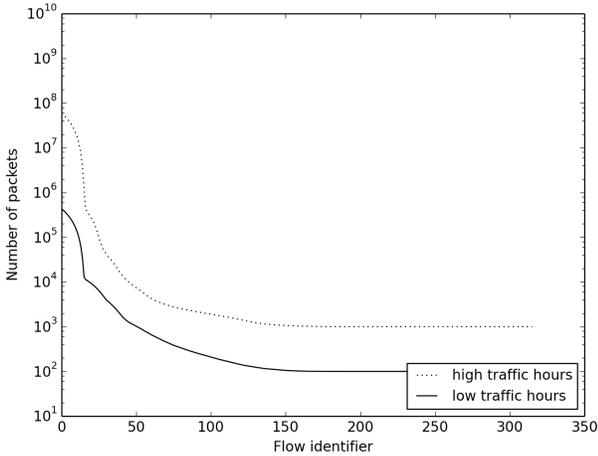

Fig. 10. Average flow size distributions of the sampled traffic as seen by the BubbleCache during high and low traffic hours.

## IV. Conclusions and Forthcoming Work

Real world network traffic presents cutoff sampling rates that can be exploited to design highly efficient elephant flow detection algorithms. In our work, we present a theoretical framework to identify these cutoff sampling rates and develop BubbleCache, a low complexity algorithm that can efficiently capture elephant flows at very high speed rates and using small resources.

In part left as future work, in this paper we did not provide performance comparisons of BubbleCache against existing solutions because we see our algorithm as a modular component that can be used to enhance any of the sampling rate base algorithms found in the literature. For instance, the well-known ElephantTrap algorithm [6] uses a static sampling rate and so it can benefit from using our Kurtosis-based algorithm to dynamically compute the traffic's optimal sampling rate avoiding any manual tuning and enabling the system to operate near the optimal tradeoff between computational scalability and accuracy.

While this paper focuses on the base theoretical aspects of the high performance detection algorithm under partial information, future work will include exploring ways in which solutions in the literature can be enhanced using our work. This will also include more performance benchmarks to compare the existing solutions with and without the enhancements of the BubbleCache's dynamic sampling rate module. In addition to these additional tests, our current work focuses also around the integration of the BubbleCache algorithm as part of a commercial software defined network (SDN) data plane to operate at port rates of 100 Gbps. We are packaging the BubbleCache in two formats: (1) as a top flow detection and ordering algorithm for SDN networks using sFlow and (2) as a high performance queue for the real time separation of elephant and mouse flows to isolate and protect them from each other.

## Appendix

*Lemma 1. Detection under partial information.* The detection likelihood of a network at time $t$ follows a multivariate hypergeometric distribution as follows:

$$P(e_\alpha(t) = 0) = P(C_\alpha(t) = F_\alpha) = \sum_{\forall \mathbf{x}' \in Z(t)} \frac{\prod_{\forall i} \binom{\sigma_i}{x'_i}}{\binom{\sum_{\forall i} \sigma_i}{\sum_{\forall i} x_i(t)}} \quad (4)$$

where $Z(t)$ is the zero quantum error region, expressed as:

$$Z_\alpha(t) = \{\mathbf{x}' \in \mathbb{N}^{|F|} \mid \sum_{\forall i} x'_i = \sum_{\forall i} x_i(t), \\ \mathbf{x}' \leq_p \boldsymbol{\sigma},\ x'_i > x'_j\ \forall i,j\ \text{s.t.}\ i \leq \alpha, j > \alpha\} \quad (5)$$

and $a \leq_p b$ means that $b$ is at least as *Pareto efficient* as $a$.

*Proof.* Assume a discrete fluid model of the network in which each flow $i$ needs to transmit a number of water droplets equal to its size metric $\sigma_i$. Flows transmit water through the network one droplet at a time and each droplet is transmitted at arbitrary times. By convention, we will assume the first and last droplets from any of the flows are transmitted at times 0 and $t_e$, respectively. An observer of the network performs only one task: counting the number of droplets each flow has transmitted and storing such information in a vector $\mathbf{x}(t)$, where each component $x_i(t)$ corresponds to the amount of droplets seen from flow $i$ up until time $t$. Based on this information, the objective is to quantify the probability that the set of flows $C_\alpha(t)$ is the same as the set of flows in $F_\alpha$.

At time $t$, the total number of droplets transmitted is $\sum_{\forall i} x_i(t)$ out of a total number of $\sum_{\forall i} \sigma_i$ droplets. The total number of possibles ways in which $\sum_{\forall i} x_i(t)$ droplets are transmitted is given by this expression:

$$\binom{\sum_{\forall i} \sigma_i}{\sum_{\forall i} x_i(t)} \qquad (6)$$

Only a subset of the total number of ways in which droplets are transmitted correspond to the case of zero quantum error. In particular, those vectors $\mathbf{x}'$ that satisfy the following conditions:

- The total number of droplets transmitted, $\sum_{\forall i} x'_i(t)$, is equal to $\sum_{\forall i} x_i(t)$.
- The number of droplets transmitted by a flow cannot be larger than its size metric: $\mathbf{x}' \leq_p \boldsymbol{\sigma}$
- The top $\alpha$ flows, $f_1, f_2, ..., f_\alpha$, are captured by the set $C_\alpha(t)$, that is, $x'_i > x'_j$ for all $i$ and $j$ such that $i \leq \alpha$ and $j > \alpha$.

The above three conditions define the zero quantum error region as expressed in Equation (5) and its cardinality is as follows:

$$|Z_\alpha(t)| = \sum_{\forall \mathbf{x}' \in Z(t)} \prod_{\forall i} \binom{\sigma_i}{x'_i} \qquad (7)$$

The probability that the quantum error is zero, $P(e_\alpha(t) = 0)$, can now be obtained from the division of Equation (7) by Equation (6).

*q.e.d.* §

*Corollary 1. Test of generality.* Assume a traffic dataset consisting of one single flow carrying $m$ packets and $n$ flows carrying 1 single packet. Then the detection likelihood function presented in Equations (4) and (5) is equivalent to Equation (2).

*Proof.* Without loss of generality, assume flow $f_1$ is the elephant flow and the rest of the flows $\{f_2, f_3, ..., f_{n+1}\}$ are the mouse flows. That is, $\sigma_1 = m$ and $\sigma_i = 1$ for all $2 \leq i \leq n+1$. Assume also that we have sampled $k$ packets from this dataset. Without loss of generality, we remove the time dimension from the following expressions.

The set $Z$ in Equation (5) can be expressed as:

$$Z = \{x' \in \mathbb{N}^{n+1} | \sum_{\forall i} x'_i = k, x_1 \leq m, x_i \leq 1, x_1 > x_i \forall i \geq 2\}$$

The denominator of Equation (4) becomes:

$$\binom{\sum_{\forall i} \sigma_i}{\sum_{\forall i} x_i} = \binom{m+n}{k}$$

We can resolve the numerator of Equation (4) using the Chu-Vandermonde identity which states the following equality [14]:

$$\sum_{x' \in Z'} \binom{\sigma_1}{x'_1} \cdots \binom{\sigma_{n+1}}{x'_{n+1}} = \binom{\sigma_1 + ... + \sigma_{n+1}}{k}$$

where $Z' = \{x' \in \mathbb{N}^{n+1} | \sum_{\forall i} x'_i = k, x_1 \leq m, x_i \leq 1 \ \forall \ i \geq 2\}$.

The above expression is almost the same as the numerator in Equation (4) except the summation is run on the set $Z'$ instead of $Z$. The main difference between these two sets is that $Z$ includes the zero quantum error constraint $x_1 > x_i, i \geq 2$. That is, $Z$ enforces that at least 2 packets have been sampled from the elephant flow so that it can be recognized from the rest of the mouse flows without incurring quantum error. Thus, we can express $Z$ as follows:

$$Z = Z' \setminus \{\{[0, x_2, ..., x_{n+1}] | \sum_{i>2} x_i = k\} \cup$$

$$\{[1, x_2, ..., x_{n+1}] | \sum_{i>2} x_i = k-1\}\}$$

That is, the set $Z$ is equal to the set $Z'$ minus the two subsets of vectors which generate quantum error. Using the above expression and the Chu-Vandermonde identity, we can express the numerator of Equation (4) as follows:

$$\sum_{\forall x' \in Z} \prod_{\forall i} \binom{\sigma_i}{x'_i} =$$

$$\sum_{x'_1 + ... + x'_{n+1} = k} \prod_{\forall i} \binom{\sigma_i}{x'_i} -$$

$$\sum_{x'_2 + ... + x'_{n+1} = k} \binom{m}{0} \prod_{\forall i \geq 2} \binom{\sigma_i}{x'_i} -$$

$$\sum_{x'_2 + ... + x'_{n+1} = k-1} \binom{m}{1} \prod_{\forall i \geq 2} \binom{\sigma_i}{x'_i} =$$

$$\binom{\sigma_1 + ... + \sigma_{n+1}}{k} - \binom{\sigma_2 + ... + \sigma_{n+1}}{k} - \binom{\sigma_2 + ... + \sigma_{n+1}}{k-1}m =$$

$$\binom{m+n}{k} - \binom{n}{k} - \binom{n}{k-1}m$$

Now dividing the above expression by the denominator we obtain the detection likelihood equation:

$$P(e_\alpha = 0)) = 1 - \frac{\binom{n}{k}}{\binom{m+n}{k}} - \frac{m\binom{n}{k-1}}{\binom{m+n}{k}}$$

The above equation is equivalent to Equation (2) for the case $2 \leq k \leq n+1$. We leave it to the reader to show that Equation (4) results in $P(e_\alpha = 0) = 1$ for the case $k > n+1$.

*q.e.d.* §

*Corollary 2. Estimated detection likelihood.* Let an elephant flow detection algorithm process traffic by sampling packets at a rate $p$. Then, an estimation of the detection likelihood can be obtained as follows:

$$\hat{P}(e_\alpha(t) = 0) = \sum_{\forall x' \in \hat{Z}(t)} \frac{\prod_{\forall i} \binom{x_i(t)/p}{x'_i}}{\binom{\sum_{\forall i} x_i(t)/p}{\sum_{\forall i} x_i(t)}} \quad (8)$$

where $\hat{Z}(t)$ is the estimated zero quantum error:

$$\hat{Z}_\alpha(t) = \{x' \in \mathbb{N}^{|F|} \mid \sum_{\forall i} x'_i = \sum_{\forall i} x_i, \\ x' \leq_p x(t)/p,\ x'_i > x'_j\ \forall i, j\ \text{s.t.}\ i \leq \alpha, j > \alpha\} \quad (9)$$

*Proof.* The expression for $\hat{P}(e_\alpha(t) = 0)$ estimates the actual detection likelihood at time $t$ by making two assumptions. First, that all packets of the same flow have been sampled at an exact rate of $p$, so that at time $t$, the actual number of transmitted packets by flow $f_i$ is $x_i(t)/p$. Second, that the current time is $t_e$, $t = t_e$, so that the actual size of each flow is $\sigma_i = x_i(t)/p$. By the law of large numbers, the first assumption is valid if the number of samples is very large. This is in general true for the case of computer networks, with high-performance switches designed to process more than 100 millions packets per second. The second assumption is also valid if we slightly adjust our definition of flow size. Instead of considering the complete historical performance of a flow to determine its size—including both past and future performance—Equation (8) relaxes this definition by only considering their past performance. This interpretation can also be explained in terms of the sources of uncertainty model described in the Section II.A in which there exist two sources of uncertainty: future uncertainty due to our limited ability to forecast the future, and past uncertainty due to voluntary packet sampling or involuntary packet dropping. While the expression of the detection likelihood in Equation (4) assumes the source of uncertainty is only due to future uncertainty, Equation (8) assumes it is only due to past uncertainty. Both equations deal with the same problem: reconstructing the size of flows for which we have only observed a partial subset of their total traffic.

*q.e.d.* §

*Lemma 2. Reconstruction under partial information.* Let $F$ be a set of flows transmitting data over a network and let $x_i$ be the size of flow $f_i$ when traffic is sampled at a rate $p$, for $0 \leq p \leq 1$. Then there exists a sampling rate $p_c < 1$ such that:

- If $x_i \gg x_j$ and $p \geq p_c$, then $\sigma_i \gg \sigma_j$ with high probability.
- If $\sigma_i \gg \sigma_j$ and $p \geq p_c$, then $x_i \gg x_j$ with high probability.

*Proof.* Consider the first statement and suppose that there exists a pair of flows $f_i$ and $f_j$ such that $x_i \gg x_j$. There are three possible cases:

- $\sigma_i > \sigma_j$ or
- $\sigma_i = \sigma_j$ or
- $\sigma_i < \sigma_j$.

Assume $\sigma_i = \sigma_j = \sigma$ and let $X_k$ be a random variable such that $X_k = x_k$ corresponds to the event "flow $f_k$ has a size $x_k$ when traffic is sampled at a rate $p$". Since $\sigma_i = \sigma_j$, then by symmetry the expected value of both $X_i$ and $X_j$ must be the same and equal to $\bar{x} = (x_i + x_j)/2$. Let $\varrho(p)$ be the probability of the event $X_i = x_i \cap X_j = x_j \cap x_i \gg x_j$ divided by the probability of the expected event $X_i = \bar{x} \cap X_j = \bar{x}$ when traffic is sampled at a rate $p$. Since $X_i = \bar{x} \cap X_j = \bar{x}$ is the expected outcome when $\sigma_i = \sigma_j$, this parameter provides a measurement of the likelihood that $\sigma_i = \sigma_j$ is true when $x_i \gg x_j$: if $\varrho(p)$ is close to zero, then $x_i \gg x_j$ is much less likely than the expected outcome, making the assumption $\sigma_i = \sigma_j$ unlikely; if $\varrho(p)$ is close to 1, then $x_i \gg x_j$ is as likely as the expected outcome, which makes $\sigma_i = \sigma_j$ possible. We have that:

$$\rho(p) = \frac{P(X_i = x_i \cap X_j = x_j \cap x_i \gg x_j)}{P(X_i = \bar{x} \cap X_j = \bar{x})} = \frac{\binom{\sigma}{x_i}\binom{\sigma}{x_j}/\binom{2\sigma}{x_i+x_j}}{\binom{\sigma}{\bar{x}}^2/\binom{2\sigma}{x_i+x_j}} =$$

$$= \frac{\frac{\sigma!}{x_i!(\sigma-x_i)!}\frac{\sigma!}{x_j!(\sigma-x_j)!}}{\left(\frac{\sigma!}{\bar{x}!(\sigma-\bar{x})!}\right)^2} = \frac{\bar{x}!}{x_i!}\frac{(\sigma-\bar{x})!}{(\sigma-x_1)!}\frac{\bar{x}!}{x_j!}\frac{(\sigma-\bar{x})!}{(\sigma-x_j)!} =$$

$$= \frac{\bar{x}\cdot\ldots\cdot(x_j+1)}{x_i\cdot\ldots\cdot(\bar{x}+1)} \times \frac{(\sigma-\bar{x})\cdot\ldots\cdot(\sigma-x_i+1)}{(\sigma-x_j)\cdot\ldots\cdot(\sigma-\bar{x}+1)} =$$

$$\rho(p) = \frac{\prod_{k=1}^{(x_i-x_j)/2}(x_j+k)}{\prod_{k=1}^{(x_i-x_j)/2}(\bar{x}+k)} \times \frac{\prod_{k=1}^{(x_i-x_j)/2}(\sigma-x_i+k)}{\prod_{k=1}^{(x_i-x_j)/2}(\sigma-\bar{x}+k)} \quad (10)$$

The following must be true:

- Since $x_i > x_j$, then $x_j < \bar{x}$ and $x_i > \bar{x}$. From Equation (10) this implies $P(X_i = x_i \cap X_j = x_j \cap x_i \gg x_j)$ is smaller than $P(X_i = \bar{x} \cap X_j = \bar{x})$, for $0 \leq p \leq 1$. Hence:

$$\rho(p) < 1 \quad for \quad 0 < p < 1 \quad (11)$$

- If $p \to 0$, then $x_i \to 0$ and $x_j \to 0$. From Equation (10) this implies

$$P(X_i = x_i \cap X_j = x_j \cap x_i \gg x_j) \to P(X_i = \bar{x} \cap X_j = \bar{x})$$

and thus:

$$\lim_{p \to 0} \rho(p) = 1 \quad (12)$$

- If $p \to 1$, then $x_i \to \sigma_i = \sigma$ and $x_j \to \sigma_j = \sigma$, which means that $P(X_i = x_i \cap X_j = x_j \cap x_i \gg x_j) \to 0$ and $P(X_i = \bar{x} \cap X_j = \bar{x}) \to 1$, leading to:

$$\lim_{p \to 1} \rho(p) = 0 \qquad (13)$$

That is: $\sigma_i = \sigma_j$ is false when $p = 1$, unlikely when $p \to 1$, and possible when $p \to 0$. Because $\varrho(p)$ is continuous with $p$, there must exist a $p_c$ such that if $p > p_c$, then $\sigma_i = \sigma_j$ is unlikely.

Now consider the case $\sigma_i < \sigma_j$. Since $x_i > x_j$, then $\sigma_i < \sigma_j$ is less likely than $\sigma_i = \sigma_j$, which means Equations (11), (12) and (13) also hold. Thus we must conclude that if $x_i > x_j$ and $p > p_c$, then $\sigma_i > \sigma_j$ with high probability.

Finally, from Equation (10), the value of $\varrho(p)$ rapidly decreases as the value $x_i - x_j$ increases. This means that the higher the value of $x_i - x_j$, the more likely the value $\sigma_i - \sigma_j$ is higher too, provided that $p > p_c$. Hence, $x_i \gg x_j$ and $p \geq p_c$ imply $\sigma_i \gg \sigma_j$ with high probability, which proves the first statement in the lemma.

The second statement in the lemma can also be demonstrated following a similar approach and is omitted from this text.

*q.e.d.* §

*Corollary 3.* Reconstruction properties under partial information. Let $F$ be a set of flows transmitting data over a network and assume the traffic dataset generated by the flows is heavy tailed according to Definition 3. Let also $x_i$ be the size of flow $f_i$ when traffic is sampled at a rate $p$, for $0 \leq p \leq 1$ and $1 \leq i \leq |F|$. Then the following is true:

(R1) There exists a cutoff sampling rate $p_c$ such that for any sampling rate $p \geq p_c$, $\sigma_i \gg \sigma_j$ implies $x_i \gg x_j$ with high probability.

(R2) The more heavy tailed the traffic data set is the lower the cutoff sampling rate $p_c$.

(R3) If the sequence $\{x_1, x_2, ..., x_{|F|}\}$ is heavy tailed, then $x_i \gg x_j$ implies $\sigma_i \gg \sigma_j$ with high probability.

(R4) If the sequence $\{x_1, x_2, ..., x_{|F|}\}$ is not heavy tailed, then either $p < p_c$ or the traffic dataset is not heavy tailed, or both.

*Proof.* R1 is a restatement of Lemma 2 applied to the case of heavy tailed traffic. R2 can be easily seen from Equation (10) and Definition 3: the more heavy tailed a traffic data set is, the larger the value of $x_i - x_j$; as a result, $\varrho(p)$ becomes smaller, which means the cutoff sampling rate also becomes smaller. R3 is also a restatement of Lemma 2. R4 is true because it is the negative form of Lemma 2 and thus can be shown by contradiction: if $p \geq p_c$ and the traffic dataset is heavy tailed, then from Lemma 2 we know $\sigma_i \gg \sigma_j$ implies $x_i \gg x_j$ with high probability; but that contradicts the assumption that $\{x_1, x_2, ..., x_{|F|}\}$ is not heavy tailed.

*q.e.d.* §